\documentclass[preprint,12pt]{elsarticle}
\usepackage{amssymb}
\usepackage{amsmath}
\usepackage{hyperref}

\journal{Physics Letters A}

\begin{document}

\begin{frontmatter}

\title{Asteroid impact, Schumann resonances and the end of dinosaurs}

\author{ Z.~K.~Silagadze}
\address{Budker Institute of Nuclear Physics and
Novosibirsk State University, 630 090, Novosibirsk, Russia}
\ead{silagadze@inp.nsk.su}

\begin{abstract}
It is believed that an asteroid/comet impact 65 million years ago ended
the dinosaur era. The researchers named the corresponding impact crater 
Chicxulub, the Mayan word roughly translated as "the dragon's tail."
We estimate the expected magnitudes of the Schumann resonance fields
immediately after the Chicxulub  impact and show that they exceed their
present-day values by about $5\times 10^4$ times. Long-term distortion
of the Schumann resonance parameters is also expected due to the 
environmental impact  of the Chicxulub event. If Schumann resonances play 
a regulatory biological role, as some studies indicate, it is possible
that the excitation and distortion of Schumann resonances after the
asteroid/comet  impact was a possible stress factor, which, among other 
stress factors associated with the impact, contributed to the demise of 
dinosaurs.
\end{abstract}


\end{frontmatter}

\section{Introduction}
Dinosaurs have been the dominant group of living organisms on the Earth for
over 160 million years (Myr). There were over 1000 species of dinosaurs
distributed worldwide. The direct evolutionary descendants of non-avian
dinosaurs, birds still make up one of the most proliferate and diverse group of
vertebrates. However, non-avian dinosaurs themselves suddenly disappeared
about 66 Myr ago \cite{1}.

There are astounding number and variety of hypotheses about causes of the
dinosaur extinction \cite{1,2}. However, the current research is concentrated
around three major ones: 1) an impact of a giant bolid (asteroid or comet)
\cite{3,4,5}; 2) Volcanic activity in modern-day India's Deccan Traps \cite{6};
3) Marine regression (drop in sea level) and the corresponding global
environmental deterioration \cite{7,8}.

All three of the above stress factors occurred at the end of the Cretaceous,
which makes it difficult to disentangle their relative importance in
the mass extinction event that occurred at the Cretaceous-Palogene
(K-Pg) boundary (formerly Cretaceous-Tertiary or K-T boundary).
It is noteworthy that  there is little evidence that bolide impacts on the 
Earth correlate well with episodes of mass extinction other than K-Pg (there 
have been five mass extinctions in the past 550 Myr),  while both sea
regression and massive volcanism do correlate well with such episodes  
\cite{2}.

Nevertheless, modern research \cite{1,9} found support for a bolide 
impact as the primary factor of the end-Cretaceous mass extinction.
Evidence of the bolid impact, coinciding in time with dinosaur extinction,
is  ubiquitous, including the huge 150 km wide Chicxulub crater in the 
Yucat\'{a}n Peninsula in Mexico, impact related iridium anomaly worldwide, 
sediments in various areas of the world dominated by impact melt spherules 
and  an unusually large amount of shocked quartz.

Yet another evidence of the enormous power of the Chicxulub impact has  
been discovered recently \cite{10}. After the impact, billions of tons of 
molten and vaporized rock was thrown  in all directions. After about ten 
minutes, these debris reached Tanis, a place at a distance of about 3 000 km 
from the impact. Bead-sized material, glassy tektites, fell from the sky, 
piercing everything in its path. Fossil fish at Tanis, densely packed in the 
deposit, are found with the impact-induced  spherules embedded in their gills.

At about the same time, strong seismic waves, generated by the Chicxulub 
impact,  arrived at Tanis generating seiche inundation surge with about 
10 m amplitude \cite{10}.

Observations at Tanis expand our knowledge of the destructive effects of  
the Chicxulub impact, identifying the potential mechanism for sudden and
extensive damage to the environment, delivered  minutes after the impact
to widely separated regions.

Our goal in this note is to show that the global extinction event 
could have had another very rapidly delivered global precursor, namely the 
excitation of Schumann resonances with currently unknown but potentially 
dangerous biological effects.

\section{Schumann resonances}
It is useful to introduce the following complex combination of the
electric and magnetic fields, called Riemann-Silberstein vector in
\cite{11}\footnote{Maybe we have here an example of the zeroth 
theorem of the history of science: a discovery (rule, regularity, insight), 
named after someone, almost always did not originate with that person
\cite{12}.  There are reasons to believe that this complex vector was first
introduced by Heinrich Weber in his 1901 book on the partial differential 
equations of mathematical physics based on Riemann's lecture notes
\cite{13}.}:
\begin{equation}
\vec{F}=\sqrt{\frac{2}{\epsilon}}\left (\frac{\vec{D}}{\sqrt{2\epsilon}}+
\frac{\vec{B}}{\sqrt{2\mu}}\right )=\vec{E}+ic\vec{B},
\label{eq1}
\end{equation}
where $\epsilon$ is the dielectric constant, $\mu$ is the magnetic 
permeability, and $c=1/\sqrt{\epsilon\mu}$ is the light velocity in a 
homogeneous and static medium in which $\vec{D}=\epsilon\vec{E}$ 
and $\vec{B}=\mu\vec{H}$.

In terms of the Riemann-Silberstein vector, the Maxwell equations read
\begin{equation}
i\frac{\partial \vec{F}}{\partial t}=c\nabla \times \vec{F},\;\;\; \nabla\cdot
\vec{F}=0.
\label{eq2}
\end{equation}
It is well known that the process of solving the Maxwell equations can be
facilitated by the use of potentials. The Riemann-Silberstein vector can
be expressed in terms of the Hertz vector (superpotential) as follows
\cite{11}:
\begin{equation}
\vec{F}=\left [\frac{i}{c}\,\frac{\partial}{\partial t}+\nabla\times\,\right ]
\left (\nabla\times\vec{\Pi}\right ).
\label{eq3}
\end{equation}
It follows from the first Maxwell equation in (\ref{eq2}) that the Hertz
superpotential $\vec{\Pi}$ must satisfy the equation
\begin{equation}
\left (\frac{1}{c^2}\,\frac{\partial^2}{\partial t^2}-\Delta\right )
\left (\nabla\times\vec{\Pi}\right )=0.
\label{eq4}
\end{equation}
The name "superpotential" indicates that electric and magnetic fields are 
expressed through the second derivatives of the superpotential, and not 
through the first derivatives, as in the case of standard potentials.

Much like more familiar four-potential, Hertz vector is not determined
uniquely by (\ref{eq3}).  In fact the group of gauge transformations of the 
Hertz vector that leave the Riemann-Silberstein vector unchanged is very 
large \cite{11,14}. In particular, when dealing with radiation fields produced
by localized sources, a very convenient choice is the assumption that the  
Hertz superpotential is radial \cite{11}:
\begin{equation}
\vec{\Pi}(\vec{r},t)=\vec{r}\,\Phi(\vec{r},t).
\label{eq5}
\end{equation}
Then it follows from (\ref{eq4}), since the operator $\hat{\vec{L}}=
\vec{r}\times\nabla$ commutes with the Laplacian, that the complex
function $\Phi(\vec{r},t)=U(\vec{r},t)+iV(\vec{r},t)$ can be adjusted in such
a way using the gauge freedom that it satisfies the wave equation
\cite{15,16}:
\begin{equation}
\left (\frac{1}{c^2}\,\frac{\partial^2}{\partial t^2}-\Delta\right )\Phi=0.
\label{eq6}
\end{equation}
Using well-known expressions of differential operators in spherical
coordinates, we get from  (\ref{eq3}) and (\ref{eq5}) 
\begin{equation}
F_r=-\frac{1}{r^2\sin{\theta}}\,\frac{\partial}{\partial\theta}\left (
\sin{\theta}\,\frac{\partial ( r\Phi)}{\partial \theta}\right )-\frac{1}
{r^2\sin^2{\theta}}\,\frac{\partial^2( r\Phi)}{\partial \varphi^2},
\label{eq7}
\end{equation}
which by using  (\ref{eq6}), and assuming harmonic dependence of fields on
time of the form $e^{\pm i\omega t}$, can be transformed into
\begin{equation}
F_r=\left [\frac{\partial^2}{\partial r^2}+\frac{\omega^2}{c^2}\right ](r\Phi).
\label{eq8}
\end{equation}
The real and imaginary parts of $\Phi$, $U$ and $V$ respectively, are called
electric and magnetic Debye (super)potentials. They are scalars under the
(proper) three-dimensional rotations, but have complicated transformation
properties under the Lorentz boosts \cite{15}.

Extreme ultraviolet and X-ray radiation from the sun ionizes atoms and 
molecules in the Earth's upper atmosphere. As a result, at altitudes of several 
tens of kilometers above the ground, the air conductivity becomes noticeable, 
and then rapidly increases by six or more orders of magnitude at heights of 
about 80 km, which indicates the beginning of the ionosphere, a region filled 
with weakly ionized gas. Thus, a spherical cavity is formed between two 
relatively good conductors: the lower one is the ground, and the upper one is 
the ionospheric plasma.

Depending on the wavelength of the electromagnetic wave, this spherical cavity 
can be considered as a waveguide, resonator or capacitor \cite{27}. At zero 
frequency (direct current) we can speak of a spherical capacitor, the upper 
"electrode" of which carries a potential of +250 kV relative to the ground. 
A corresponding electrostatic field, the so-called fair weather field of about 
100 V/m at the ground, is observed worldwide in local quiet weather conditions.

For radio frequencies with wavelengths much smaller than the Earth's radius, 
the cavity acts as a waveguide up to frequencies of 10-20 MHz, above which 
the ionosphere becomes transparent.

Finally, for very low frequency electromagnetic waves with wavelengths 
comparable to the circumference of the Earth, the Earth-ionosphere cavity 
can be considered as an electromagnetic resonator, and the corresponding 
global electromagnetic resonances are called Schumann resonances \cite{27}.

More precisely, Schumann resonances are quasi-standing transverse magnetic 
modes in the Earth-ionosphere cavity, in which the radial component of the 
magnetic field equals to zero and thus (\ref{eq8}) implies   $V=0$. Then from 
(\ref{eq3})
\begin{equation}
\vec{E}=\nabla\times\left (\nabla\times\vec{r}\right )U,\;\;\;
\vec{B}=\frac{1}{c^2}\,\frac{\partial}{\partial t}\,\nabla\times\vec{r}\,U,
\label{eq9}
\end{equation}
and for the harmonic (complex) fields with $e^{i\omega t}$ time dependence
we get 
\begin{eqnarray} &&
E_r=\left [\frac{\partial^2}{\partial r^2}+\frac{\omega^2}{c^2}\right ](rU),\;
E_\theta=\frac{1}{r}\,\frac{\partial^2(rU)}{\partial r\partial\theta},\;
E_\varphi=\frac{1}{r\sin{\theta}}\,\frac{\partial^2(rU)}{\partial r
\partial\varphi},\nonumber \\ &&
B_\theta=\frac{i\omega}{c^2}\,\frac{1}{\sin{\theta}}\,
\frac{\partial U}{\partial\varphi},\;B_\varphi=-\frac{i\omega}{c^2}\,
\frac{\partial U}{\partial\theta}.
\label{eq10}
\end{eqnarray}
The wave equation (\ref{eq6}) for U (with $e^{i\omega t}$ harmonic time 
dependence) can be solved by separation of variables in spherical
coordinates. Namely, taking $U(\vec{r},t)=\rho(r)\,Y_{lm}(\theta,\varphi)
\,e^{i\omega t}$, where $Y_{lm}(\theta,\varphi)$ are spherical functions, 
and using
\begin{equation}
\Delta=\frac{1}{r^2}\,\frac{\partial}{\partial r}\left (r^2\frac{\partial}
{\partial r}\right )+\frac{1}{r^2}\,\Delta_\perp,
\label{eq11}
\end{equation}
where $\Delta_\perp$ is the angular part of the Laplacian with
$\Delta_\perp Y_{lm}=-l(l+1) Y_{lm}$, we get the spherical Bessel 
differential equation for the radial function $\rho(r)$:
\begin{equation}
\left [\frac{d^2}{dr^2}+\frac{2}{r}\,\frac{d}{dr}+k^2-\frac{l(l+1)}{r^2}
\right ]\rho(r)=0,\;\;\;k=\frac{\omega}{c}.
\label{eq12}
\end{equation}
Therefore, inside the Earth-ionosphere cavity the Fourier component of
$U$ has the form
\begin{equation}
U(\vec{r},\omega)=\sum\limits_{l=0}^\infty\sum\limits_{m=-l}^l\left [A_{lm}\,h_l^{(1)}
(kr)+B_{lm}\,h_l^{(2)}(kr)\right ]Y_{lm}(\theta,\varphi),
\label{eq13}
\end{equation}
where $h_l^{(1)}(kr)$ and $h_l^{(2)}(kr)$ are spherical Hankel functions
of the first and second kinds, respectively. Since we have chosen 
$e^{i\omega t}$ for our time evolution,  $h_l^{(1)}(kr)$ corresponds to
the spherical incoming wave, while $h_l^{(2)}(kr)$ --- to the spherical
outgoing wave.

A real  Earth-ionosphere waveguide have a very complicated 
configuration. Here we assume a simplified model \cite{17}.
Earth is considered as a perfectly conducting sphere of radius $R$.
It is further assumed that the ionosphere begins with an inner radius 
of $R+h$ and is an infinite, uniform and isotropic plasma with complex 
dielectric constant (the imaginary part of which is proportional to the 
plasma conductivity \cite{18}).

In the ionosphere we can have only outgoing spherical waves (Sommerfeld
radiation condition  \cite{19}). Thus
\begin{equation}
U(\vec{r},\omega)=\sum\limits_{l=0}^\infty\sum\limits_{m=-l}^l C_{lm}\,h_l^{(2)}
(kr)\,Y_{lm}(\theta,\varphi)\,e^{i\omega t}, \;\;\; r>R+h.
\label{eq14}
\end{equation}
Schumann resonance frequencies are determined by boundary conditions
at $r=R$ and $r=R+h$ \cite{17} (for somewhat different approach, see
\cite{20}). Namely, at $r=R$ the tangential components of the electric field
must vanish. This leads to the condition
\begin{equation}
\left . \frac{\partial}{\partial r}(rU)\right |_{r=R}=0.
\label{eq15}
\end{equation}
At $r=R+h$, the tangential components of the electric field $\vec{E}$, and
the tangential components of the magnetic field $\vec{H}=\vec{B}/\mu$
must be continuous. If in the cavity $\epsilon\approx\epsilon_0$,
$\mu\approx\mu_0$, while in the ionosphere $\epsilon=\hat{\epsilon}\,
\epsilon_0$, $\mu\approx\mu_0$,  in light of (\ref{eq10}), the continuity
conditions take the form
\begin{eqnarray} &&
\left . \frac{\partial}{\partial r}(rU)\right |_{r=(R+h)_-}=
\left . \frac{\partial}{\partial r}(rU)\right |_{r=(R+h)_+},\nonumber \\ &&
U(r=(R+h)_-)=\hat{\epsilon}\,U(r=(R+h)_+).
\label{eq16}
\end{eqnarray}
If we substitute (\ref{eq13}) and  (\ref{eq14}) into (\ref{eq15}) and  
(\ref{eq16}), we get a homogeneous system of linear equations
\begin{eqnarray} &
u_l^\prime(kR)\,A_{lm}+v_l^\prime(kR)\,B_{lm}=0, & \nonumber \\ &
u_l^\prime(k(R+h))\,A_{lm}+v_l^\prime(k(R+h))\,B_{lm}-
v_l^\prime(\sqrt{\hat{\epsilon}}\,k(R+h))\,C_{lm}=0,  &\nonumber \\ &
u_l(k(R+h))\,A_{lm}+v_l(k(R+h))\,B_{lm}-\sqrt{\hat{\epsilon}}\,
v_l(\sqrt{\hat{\epsilon}}\,k(R+h))\,C_{lm}=0. \;\;& 
\label{eq17}
\end{eqnarray}
Here we  introduced the notations \cite{17}
\begin{equation}
u_l(x)=x h_l^{(1)}(x),\;v_l(x)=x h_l^{(2)}(x),\;
u_l^\prime(x)=\frac{d u_l(x)}{dx},\;v_l^\prime(x)=\frac{d v_l(x)}{dx}.
\label{eq18}
\end{equation}
The system (\ref{eq17}) has a non-trivial solution for $A_{lm}$,  $B_{lm}$,  
$C_{lm}$, only when the $3\times 3$ determinant of its coefficients equals 
zero. This requirement yields the following equation  \cite{17}
\begin{eqnarray} &&
u_l^\prime(kR)\,v_l^\prime(k(R+h))-u_l^\prime(k(R+h))\,v_l^\prime(kR)=
\nonumber \\ &&
\frac{1}{\sqrt{\hat{\epsilon}}}\frac{v_l^\prime\left (\sqrt{\hat{\epsilon}}\,
k(R+h)\right )}{v_l\left (\sqrt{\hat{\epsilon}}\,k(R+h)\right )}\left [ 
u_l^\prime(kR)\,v_l(k(R+h))-u_l(k(R+h))\,v_l^\prime(kR)\right ]. \;\;\;\;
\label{eq19}
\end{eqnarray}
The solutions of this transcendental equation with respect to $k$ determine
eigenmodes $\omega=c_0\,k$ of the Earth-ionosphere resonator, which are 
called Schumann resonances. Here $c_0$ is the speed of light inside the 
resonator, which is the same as the light velocity in vacuum, 
$c_0=1/\sqrt{\epsilon_0\mu_0}$, for the approximations used. In general, 
$\omega$ is a complex number. Its real part  gives the eigenfrequency of the 
resonator, while the imaginary part determines the resonance width, since it 
corresponds to the damping factor of the eigenmode. The resonance width is 
characterized by the quality factor $Q=\omega/\Delta \omega$, where 
$\Delta\omega$ is the resonance width at half maximum.

In the crude approximation of the infinite conductivity of the ionosphere, the
right-hand-side of (\ref{eq19}) vanishes and the equation for the 
eigenfrequencies simplifies. Further simplification can be achieved by using
$h\ll R$,  so that we can expand $u_l^\prime(k(R+h))\approx u_l^\prime(kR)
+ u_l^{\prime\prime}(kR)\,kh$, and $v_l^\prime(k(R+h))\approx v_l^\prime(kR)
+ v_l^{\prime\prime}(kR)\,kh$. Besides, it follows from the definitions of
$u_l(x)$ and $v_l(x)$, that
\begin{eqnarray} &&
u_l^\prime(x)=h_l^{(1)}(x)+x\frac{d h_l^{(1)}(x)}{dx},\;\;
u_l^{\prime\prime}(x)=2\frac{h_l^{(1)}(x)}{dx}+x\frac{d^2 h_l^{(1)}(x)}{dx^2},
\nonumber \\ &&
v_l^\prime(x)=h_l^{(2)}(x)+x\frac{d h_l^{(2)}(x)}{dx},\;\;
v_l^{\prime\prime}(x)=2\frac{h_l^{(2)}(x)}{dx}+x\frac{d^2 h_l^{(2)}(x)}{dx^2}.
\label{eq20}
\end{eqnarray}
The second derivatives of the Hankel functions can be eliminated by using
the fact that $\rho(r)=h_l^{(1)}(kr)$ and  $\rho(r)=h_l^{(2)}(kr)$ satisfy the
differential equation (\ref{eq12}). This gives
\begin{equation}
u_l^{\prime\prime}(x)=\left (\frac{l(l+1)}{x^2}-1\right )xh_l^{(1)}(x),\;\;\;
v_l^{\prime\prime}(x)=\left (\frac{l(l+1)}{x^2}-1\right )xh_l^{(2)}(x).
\label{eq21}
\end{equation}
Taking all this into account, in the case of perfectly conducting ionosphere,
equation (\ref{eq19}) simplifies to
\begin{equation}
\left (1-\frac{l(l+1)}{k^2R^2}\right )\left [h_l^{(1)}(x)\frac{d h_l^{(2)}(x)}{dx}-
h_l^{(2)}(x)\frac{d h_l^{(1)}(x)}{dx}\right ]_{x=kR}=0.
\label{eq22}
\end{equation}
The expression in the square brackets is the Wronskian of $h_l^{(1)}(x)$
and $h_l^{(2)}(x)$, and it is not zero, because these two solutions
of the spherical Bessel equations are independent. Therefore, from
(\ref{eq22}) we get $\omega_l=c_0k_l=\frac{c_0}{R}\sqrt{l(l+1)}$, and in this
crude approximation, the Schumann resonance frequencies are
\begin{equation}
f_l=\frac{\omega_l}{2\pi}=\frac{c_0}{2\pi R}\sqrt{l(l+1)},\;\;l=1,2,\ldots
\label{eq23}
\end{equation}
The observed frequencies of the first five Schumann resonances are 7.8,
14.1, 20.3, 26.4 and 32.5 Hz, respectively \cite{21}, and they are about
25\% lower than it follows from (\ref{eq23}).  This discrepancy is due to 
neglecting the effects of electric losses on the resonance frequencies. 

When considering Schumann resonances, there are two sources of losses 
present \cite{21A}. There are losses due to finite  electrical conductivity 
of the atmosphere that increases with altitude, and there are losses due 
to the fact that both the Earth's surface and the ionosphere are good but 
imperfect conductors. As is known from a damped harmonic oscillator, the 
inclusion  of the damping effects not only decreases amplitude over time, 
but also reduces the resonant frequency.

However, more realistic models of ionosphere lead to far more complex
theories that in general defy analytical treatment. The  ionosphere structures
considered include two-layer and multi-layer models, two-exponential, ``knee'' 
and ``multi-knee'' profiles (for relevant references see \cite{23}).

Let us take a quick look at the two-scale-height model of the ionosphere that 
captures the essential underlying physics of a system with conductivity that 
increases with height \cite{21B,21C,21D}.

The Joule dissipation responsible for the damping is, in a good approximation, 
confined in height to two layers. The first, the so-called conduction boundary 
at the altitude $h_1$ is defined by the condition of equality of the  
displacement and conduction currents, $\sigma_1=\epsilon_0\omega$, so that at 
several scale lengths ($\xi_1$) below $h_1$ the atmosphere is insulating, and
above several scale lengths it is conducting. 

The second,  the so-called reflection boundary at $h_2$ is the altitude at
which $\omega\tau_D\approx 1$, where $\tau_D$ is the magnetic diffusion time 
through a conductivity  scale length $\xi_2$ at $h_2$ \cite{21C}. In other
words,  at $h_2$ the character of the fields changes from being wavelike to 
diffusion-like \cite{21D}. In terms of conductivity $\sigma_2$ at $h_2$, this
condition has the form $4\mu_0\omega\sigma_2\xi^2_2=1$ \cite{21C}.    

The conductivity profiles in these two characteristic layers of ionosphere are
given by exponential functions with different scale lengths $\xi_1$ and 
$\xi_2$: $\sigma(z)=\sigma_1\exp{[(z-h_1)/\xi_1]}$ and $\sigma(z)=\sigma_2
\exp{[(z-h_2)/\xi_2]}$ respectively. It can be shown \cite{21B,21D} that for 
such conductivity profiles the real parts of the Schumann resonance 
eigenfrequencies and the Q-factors are given by
\begin{equation}
f_l\approx\frac{c_0}{2\pi R}\sqrt{l(l+1)\frac{h_1}{h_2}},\;\;
Q_l\approx \frac{2}{\pi}\left (\frac{\xi_1}{h_1}+\frac{\xi_2}{h_2}\right)^{-1}.
\label{eq23A}
\end{equation}
Using typical values of the two-scale-height model $h_1=50~\mathrm{km}$,
$h_1=90~\mathrm{km}$, $\xi_1=\xi_2=5~\mathrm{km}$, these expressions yield
results  that are in quite good agreement with observations \cite{21B}. 

Winfried Otto Schumann, a professor at the Technische Hochschule 
M\"{u}n\-chen, rightfully gets most of the credit for predicting Schumann 
Resonances. However, Schumann resonance history is an interesting story 
\cite{22}. The idea of natural global electromagnetic resonances goes back
to George F. Fitzgerald in 1893 and Nikola Tesla in 1905 \cite{22,23}. The
formula (\ref{eq23}) for resonance frequencies of a spherical condenser was
first obtained by Joseph Larmor already in 1894 \cite{22}.

Above we have outlined just some basics of Schumann resonances for the
reader's convenience. More detailed information  about Schumann resonance
research can be found in  books \cite{17,24,25,26,27}.

\section{Excitation of Schumann resonances by an asteroid impact}
Schumann resonances are excited primarily by lightning discharges. On the 
other hand, it is known that explosions and hypervelocity impacts are
accompanied by macroscopic charge separation \cite{28,29,30}. Upon 
a hypervelocity impact, a partially ionized plasma is formed, which rapidly
expands. In addition to plasma, the impact will result in the formation of 
molten and fragmented debris of the target material, which are expected to 
become negatively charged when in contact with the plasma, since electrons 
are much more mobile than ions. The subsequent inertial separation of the 
positively charged plasma and the negatively charged debris will lead to the 
separation of charge over macroscopic distances \cite{30}.

One can also imagine some other mechanisms of charge separation, for example, 
those that act during the dust storms \cite{31} and volcanic eruptions 
\cite{32}. Therefore, we assume that an asteroid impact is immediately 
accompanied by a thunderstorm with a large number of lightning discharges. 
Namely, let $dN(t)=Ne^{-t/T}\frac{dt}{T}$ be the number of lightning discharges 
during a time period $dt$ at the time $t$ after the impact. Here 
$T\approx 100~\mathrm{s}$ is the transient crater formation time for the 
Chicxulub event \cite{ 33}, and $N$ is the total number of lightning strikes 
in the impact thunderstorm. If the current in an average individual lightning 
strike is $I_0e^{-t/\tau}$, with $\tau\approx 500~\mu\mathrm{s}$ and 
$I_0\approx 2\cdot 10^4~\mathrm{A}$ \cite{34}, the total current will be
\begin{equation}
I(t)=\int\limits_0^t e^{-\frac{t-s}{\tau}}dN(s)=\frac{NI_0}{T}e^{-\frac{t}{\tau}}
\int\limits_0^t e^{\left (\frac{1}{\tau}-\frac{1}{T}\right )s} ds \approx
\frac{NI_0\tau}{T} e^{-\frac{t}{T}},
\label{eq24}
\end{equation}
where at the last step we have taken into account that $T\gg \tau$ (in fact,
this condition is not sufficient do discard the second exponent $e^{-t/\tau}$,
which occurs after the integration in (\ref{eq24}), since short signals with
$\omega_n\tau\sim 1$  can excite Schumann resonances just as effectively
as long signals with $\omega_n T\gg 1$. However, it will be clear from the 
final answer that we can still neglect the contribution of this term due to the 
condition $\omega_n\tau\ll1$, which is satisfied by the first few Schumann 
resonances).

Accordingly, as the current density, which we will consider having only 
a radial component, we take
\begin{equation}
j_r(\vec{r},t)=\frac{I(t)\Delta l}{2\pi r^2\sin{\theta}}\,\delta(\theta)
\delta(r-R)\Theta(t),
\label{eq25}
\end{equation}
where $\Theta(t)$ is the Heaviside step function introduced to indicate 
that there is no current for $t<0$, and $\Delta l\approx 10^3~\mathrm{m}$
\cite{34} is the length of an average lightning channel. The current density
(\ref{eq25}), when integrated over the whole space, gives the total current
moment: $\int j_r(\vec{r},t)\,dV=I(t)\Delta l$.

Now we will consider how the Schumann resonances are excited by 
the vertical electric dipole with current density (\ref{eq25}). We will closely 
follow \cite{24}, for other approaches see \cite{26} and \cite{17,34}.

Maxwell equations
\begin{equation}
\nabla\times\vec{E}=-\mu_0\,\frac{\partial \vec{H}}{\partial t},\;\;
\nabla\times\vec{H}=\epsilon_0\,\frac{\partial \vec{E}}{\partial t}+
\vec{j}(\vec{r},t),
\label{eq26}
\end{equation}
for Fourier components with $e^{i\omega t}$ time dependence take 
the form
\begin{equation}
\nabla\times\vec{E}(\vec{r},\omega)=-i\omega\mu_0\vec{H}(\vec{r},\omega),\;\;
\nabla\times\vec{H}(\vec{r},\omega)=i\omega\epsilon_0\vec{E}(\vec{r},\omega)+
\vec{j}(\vec{r},\omega),
\label{eq27}
\end{equation}
where $\vec{j}(\vec{r},\omega)$ has only the radial component
\begin{equation}
j_r(\vec{r},\omega)=\int\limits_{-\infty}^\infty e^{-i\omega t} j_r(\vec{r},t)\,dt=
\frac{NI_0\Delta l\,\tau}{1+i\omega T}\,\frac{\delta(\theta)\delta(r-R)}
{2\pi r^2\sin{\theta}}.
\label{eq28}
\end{equation}
A vertical electric dipole source at $\theta=0$ can excite only fields 
that do not have a $\varphi$-dependence. This follows from  the
azimuthal symmetry of the problem. Then it can be checked in
spherical coordinates that the fields given by equations (\ref{eq10})
still satisfy the Maxwell equations (\ref{eq27}) if  the Debye 
superpotential $U(r,\theta)$ satisfies the equation
\begin{equation}
r(\Delta+k^2)U=\left (\frac{\partial^2}{\partial r^2}+k^2\right)(rU)+
\frac{1}{r\sin{\theta}}\,\frac{\partial}{\partial \theta}\left (\sin{\theta}
\frac{\partial U}{\partial \theta}\right )=-\frac{j_r(\vec{r},\omega)}
{i\omega\epsilon_0}.
\label{eq29}
\end{equation}
Since $j_r(\vec{r},\omega)$ is proportional to $\delta(r-R)$, it vanishes 
in the Earth-ionosphere cavity. Thus, in the cavity $U(\vec{r},\omega)$
is still given by (\ref{eq13}) with the difference that only $m=0$ modes
are excited due to azimuthal symmetry, and, therefore $Y_{lm}$ spherical 
functions can be replaced simply by Legendre polynomials $P_l(\cos{\theta})$:
\begin{equation}
U(\vec{r},\omega)=\sum\limits_{n=0}^\infty \left [A_nh^{(1)}_n(kr)+
B_nh^{(2)}_n(kr)\right ]P_n(\cos{\theta}).
\label{eq30}
\end{equation}
The boundary condition at $r=R+h$ is
\begin{equation}
\left . \frac{\partial}{\partial r}(rU)\right |_{r=R+h}=0,
\label{eq31}
\end{equation} 
if the ideally conducting ionosphere is assumed. To get the boundary condition
at $r=R$, we integrate (\ref{eq29}) over $r$ from $R-\varepsilon$ to 
$R+\varepsilon$, take into account that inside the ideally conducting Earth
there is no tangential electric field and hence $\left . \frac{\partial}
{\partial r}(rU)\right |_{r=R-\varepsilon}=0$, and finally take the limit 
$\varepsilon\to 0$. As a result, we get \cite{24}
\begin{equation}
\left . \frac{\partial}{\partial r}(rU)\right |_{r=R}=-\frac{NI_0\Delta l\,\tau}
{1+i\omega T}\,\frac{\delta(\theta)}{2\pi i\epsilon_0\omega R^2\sin{\theta}}=
-\sum\limits_{n=0}^\infty a_n\,P_n(\cos{\theta}),
\label{eq32}
\end{equation} 
where 
\begin{equation}
a_n=\frac{NI_0\Delta l\,\tau}{1+i\omega T}\,\frac{1}{2\pi i\epsilon_0\omega R^2}
\left(n+\frac{1}{2}\right ),
\label{eq33}
\end{equation}
and at the last step, we have expanded $\delta(\theta)/\sin{\theta}$ into 
a series of Legendre polynomials:
\begin{equation}
\frac{\delta(\theta)}{\sin{\theta}}=\sum\limits_{n=0}^\infty \left(n+
\frac{1}{2}\right )P_n(\cos{\theta}).
\label{eq34}
\end{equation}
Substituting (\ref{eq30}) into (\ref{eq31}) and (\ref{eq32}), we get the 
following system of linear equations for unknown coefficients $A_n$ and  $B_n$:
\begin{eqnarray} &&
A_n u^\prime_n\left (k(R+h)\right )+B_n v^\prime_n\left (k(R+h)\right )=0, 
\nonumber \\ &&
A_n u^\prime_n\left (kR \right )+B_n v^\prime_n\left (kR\right )=-a_n. 
\label{eq35}
\end{eqnarray}
This system is easily solved, and if the results are substituted in 
(\ref{eq30}), we obtain
\begin{equation}
U=\sum\limits_{n=0}^\infty a_n\,\frac{v^\prime_n\left (k(R+h)
\right )h^{(1)}_n(kr)-u^\prime_n\left (k(R+h)\right )h^{(2)}_n(kr)}{
u^\prime_n\left (k(R+h)\right )v_n^\prime(kR)-v^\prime_n\left (k(R+h)\right )
u_n^\prime(kR)}\,P_n(\cos{\theta}).
\label{eq36}
\end{equation}
From (\ref{eq12}) It follows that the Hankel functions satisfy the relation
\begin{equation}
\left (\frac{d^2}{dr^2}+k^2\right )\left (rh_n^{(1,2)}(kr)\right )=
\frac{n(n+1)}{r}\,h_n^{(1,2)}(kr).
\label{eq37}
\end{equation}
Then from (\ref{eq10}) and (\ref{eq36}) we obtain the following expression 
for the Fourier component $E_r(\vec{r},\omega)$ of the electric field on the 
ground (at $r=R$):
\begin{equation}
E_r(\vec{r},\omega)=\sum\limits_{n=0}^\infty \frac{a_n n(n+1)}{R}\,
c_n\,P_n(\cos{\theta}),
\label{eq38}
\end{equation}
where
\begin{equation}
c_n=\frac{v^\prime_n\left (k(R+h)
\right )h^{(1)}_n(kR)-u^\prime_n\left (k(R+h)\right )h^{(2)}_n(kR)}{
u^\prime_n\left (k(R+h)\right )v_n^\prime(kR)-v^\prime_n\left (k(R+h)\right )
u_n^\prime(kR)}.
\label{eq39}
\end{equation}
Now we use, as in the previous section, smallness of the ratio $h/R$ and
expand both the numerator and denominator of $c_n$ in terms of this small 
quantity.  To first order, we have 
\begin{eqnarray} &&
\hspace*{-5mm}
u^\prime_n\left (k(R+h)\right )v_n^\prime(kR)-v^\prime_n\left (k(R+h)\right )p
u_n^\prime(kR)\approx kh\left (n(n+1)-k^2R^2\right )W,
\nonumber \\ &&
\hspace*{-5mm}
v^\prime_n\left (k(R+h)\right )h^{(1)}_n(kR)-u^\prime_n\left (k(R+h)\right )
h^{(2)}_n(kR)\approx kR\,W,
\label{eq40}
\end{eqnarray}
where $W$ is the Wronskian of $h_l^{(1)}(x)$ and $h_l^{(2)}(x)$ at $x=kR$.
Replacing $n(n+1)$ by $\frac{R^2}{c^2}\,\omega_n^2$, and $k^2$ by 
$\frac{\omega^2}{c^2}$, we get
\begin{eqnarray} &
E_r(\vec{r},\omega)=\sum\limits_{n=0}^\infty\frac{a_n\omega_n^2P_n(\cos{\theta})}
{h(\omega_n^2-\omega^2)}= 
\frac{NI_0\Delta l\,\tau}{2\pi i\epsilon_0\omega 
R^2h(1+i\omega T)}\sum\limits_{n=0}^\infty\frac{\omega_n^2(n+\frac{1}{2})
P_n(\cos{\theta})}{\omega_n^2-\omega^2}. &
\label{eq41}
\end{eqnarray} 
But $\frac{\omega_n^2}{\omega_n^2-\omega^2}=1+\frac{\omega^2}{\omega_n^2-
\omega^2}$, and the first term according to (\ref{eq34}) will lead to a 
$\delta(\theta)$ proportional contribution that is equal to zero outside the 
source. Therefore, finally we can write 
\begin{equation}
E_r(\vec{r},\omega)=\frac{NI_0\Delta l\,\tau}{4\pi i\epsilon_0 R^2h
(1+i\omega T)}\sum\limits_{n=0}^\infty\frac{\omega}{\omega_n^2-\omega^2}
(2n+1)P_n(\cos{\theta}).
\label{eq42}
\end{equation} 
To find the electric field in the time domain, we perform the inverse Fourier
transform of the frequency domain field $E_r(\vec{r},\omega)$ (since it is 
assumed that the Earth is perfectly conductive, the electric field on the 
ground is radial, so we omit the lower index indicating the radial component 
in $E(\vec{r},t)$):
\begin{equation}
E(\vec{r},t)=\frac{1}{2\pi}\int\limits_{-\infty}^\infty e^{i\omega t} E_r(\vec{r},
\omega)\,d\omega.
\label{eq43}
\end{equation} 
However, for the integral (\ref{eq43}) to have a well-defined meaning, it is 
necessary to indicate how to handle  the singularities of the integrand: as is
clear from  (\ref{eq42}), we have three simple poles at $\pm\omega_n$ and
$\frac{i}{T}$, and the first two of them lie on the integration contour of  
(\ref{eq43}).

This problem is solved by noting that in reality the Earth and ionosphere are
not ideal conductors and as a result the Schumann eigenfrequencies become
complex with small imaginary parts $\gamma_n=\frac{\omega_n}{2Q_n}\ll
\omega_n$ \cite{17} (the quality factors for the first Schumann resonances
are $Q_1\approx 4.63$,  $Q_2\approx 5.96$,  $Q_3\approx 6.56$,  
$Q_4\approx 6.83$,  $Q_5\approx 6.95$  \cite{17}). For a dissipative 
ionosphere, the imaginary part $\gamma_n$ of the positive pole at 
$\omega=\omega_n$ is positive. The imaginary part of the negative pole  at
$\omega=-\omega_n$ is fixed by the condition $E_r^*(\vec{r},\omega)=
E_r(\vec{r},-\omega)$ (the reality condition for the time domain field 
$E_r(\vec{r},t)$) and turns out to be also  $\gamma_n$. Therefore, we replace 
$(\omega^2-\omega_n^2)^{-1}$ in the integral (\ref{eq43}) by $[(\omega-
\omega_n-i\gamma_n)(\omega+\omega_n-i\gamma_n)]^{-1}$, close the 
integration contour in the upper half-plane where the integrand decreases 
exponentially, and evaluate the integral according to the Cauchy residue 
theorem as a sum of residues at three simple poles. As a result, we obtain
\begin{eqnarray} &
\hspace*{-7mm} E(\vec{r},t)= 
\frac{NI_0\Delta l\,\tau}{4\pi\epsilon_0R^2h}\sum\limits_{n=0}
^\infty\frac{(2n+1)P_n(\cos{\theta})}{1+\omega_n^2T^2}\left [e^{-\frac{t}{T}}-
e^{-\gamma_n t}\left (\cos{\omega_n t}+\omega_nT \sin{\omega_n t}
\right )\right ]. &
\label{eq44}
\end{eqnarray} 
The first $e^{-t/T}$ term in square braces expresses the direct, non-resonant
contribution to the electric field  from the source, while the remaining terms
correspond to the excitation of resonant  modes of the cavity \cite{34}.
Since $\omega_n T\gg 1$, the resonant part of the electric field takes the 
form
\begin{equation}
E_{res}(\vec{r},t)=-\frac{NI_0\Delta l\,\tau}{4\pi\epsilon_0R^2h}\sum
\limits_{n=0}(2n+1)\,\frac{e^{-\gamma_n t}}{\omega_n T}\sin{\omega_n t}
\,P_n(\cos{\theta}).
\label{eq45}
\end{equation} 
To estimate an average amplitude of the excitation, we replace 
$e^{-\gamma_n t}$ by its average value $\frac{1}{T}\int\limits_0^T
e^{-\gamma_n t}dt\approx \frac{1}{\gamma_n T}$ \cite{34}. In this way, 
we get for the amplitude of the first Schumann resonance
\begin{equation}
{\cal A}_1\approx \frac{3NI_0\Delta l\,\tau}{4\pi\epsilon_0R^2h
\omega_1\gamma_1T^2}=\frac{3\Delta Q\,\Delta l}{4\pi\epsilon_0R^2h
\omega_1\gamma_1T^2},
\label{eq46}
\end{equation}
where $\Delta Q=NI_0\tau$ is the total amount of electric charge separated 
by a macroscopic distance. In \cite{30} the following empirical relation was
obtained for $\Delta Q$ in laboratory scale hypervelocity impacts (all 
quantities are in the SI units)
\begin{equation}
\Delta Q\approx 10^{-2}\,m\left(\frac{V}{3000}\right )^{2.6\pm 0.1},
\label{eq47}
\end{equation}
where $m$ is the impactor mass, and $V$ is its velocity. It was argued 
\cite{35} that the Chicxulub impactor was a fast asteroid or a long-period 
comet with energy between $1.3\times 10^{24}~\mathrm{J}$ and $5.8\times 10^{25}
~\mathrm{J}$, and mass between $1.0\times 10^{15}~\mathrm{kg}$ and 
$4.6\times 10^{17}~\mathrm{kg}$. Taking the lowest numbers $m=1.0\times 
10^{15}~\mathrm{kg}$ and $E_{kin}=1.3\times 10^{24}~\mathrm{J}$, for the 
velocity we obtain $V=\sqrt{\frac{2E_{kin}}{m}}\approx 50~\mathrm{km}/
\mathrm{s}$. Then an interpolation of empirical relation (\ref{eq47}) to this 
enormous scale gives a huge number $\Delta Q\approx 1.5\times 10^{16}~
\mathrm{C}$. However, a recent simulation \cite{36} resulted in the Chicxulub 
scale impact-generated magnetic field that was three orders of magnitude 
smaller than expected from the relation (\ref{eq47}). Therefore, as a more 
realistic estimate, we will take  $\Delta Q\approx 1.5\times 10^{13}
~\mathrm{C}$. As for other parameters in (\ref{eq46}), we will assume  
$R=6400~\mathrm{km}$, $h=75~\mathrm{km}$, $\omega_1=49$ and $\gamma_1=5.3$. 
Then we get from (\ref{eq46}) the following amplitudes for the electric and 
magnetic fields of the first Schumann resonance:
\begin{equation}
{\cal A}_1\approx 50~\mathrm{V}/\mathrm{m},\;\;\; {\cal B}_1=\frac{{\cal A}_1}
{V_{ph}}\approx 230~\mathrm{nT},
\label{eq48}
\end{equation}
where $V_{ph}\approx 0.7c_0$ is the phase velocity of the electromagnetic waves 
in the earth-ionosphere cavity. For comparison, the measured Schumann 
resonance background fields are very small, of the order of $\mathrm{mV}/
\mathrm{m}$ for the electric field, and several $\mathrm{pT}$ for the magnetic
field \cite{23}. As we see, estimated magnitudes of the Chicxulub impact 
induced Schumann resonance fields exceed to their present-day values about 
$5\times 10^4$ times.

\section{On biological effects of ELF electromagnetic fields}
Shortly after Schumann and his graduate student K\"{o}nig made their first 
attempts to detect Schumann resonances, K\"{o}nig and Ankerm\"{u}ller noted 
a striking similarity between these signals and human brain 
electroencephalograms (EEG) \cite{37}.  

The classical EEG rhythms are delta (1-3 Hz), theta (4-7 Hz), alpha (8-13 Hz), 
beta (14-29 Hz) and gamma (30-80+ Hz) \cite{38}, and we can try to roughly
estimate these fundamental brain frequencies as follows \cite{39}. 

Human neocortex,  which form most of the white matter, contains about $10^{10}$ 
interconnected neurons. Imagine that the wrinkled surface of each hemisphere, 
where these neurons are situated, is inflated so that to create a spherical 
shell with effective radius $a=\sqrt{S/4\pi}$, where 
$S=1000-1500~\mathrm{cm}^2$ is the surface area of the hemisphere. 
Characteristic corticocortical axon excitation
propagation speed is $V=600-900~\mathrm{cm}/\mathrm{s}$. Therefore we can 
write the wave equation for the propagation of these excitation waves on the
surface of the sphere as follows:
\begin{equation}
\Delta \Phi(\theta,\varphi,t)=\frac{1}{V^2}\frac{\partial^2 \Phi(\theta,
\varphi,t)}{\partial t^2},
\label{eq49}
\end{equation}
where $\Phi$ is some quantity  characterizing the excitation. Because of 
spherical symmetry, we seek the solution of (\ref{eq49}) in the form
\begin{equation}
\Phi(\theta,\varphi,t)=\sum\limits_{l=0}^\infty\sum\limits_{m=-l}^l A_{lm}F_l(t)
Y_{lm}(\theta,\varphi).
\label{eq50}
\end{equation}
Recalling (\ref{eq11}) and taking into account that $r=a=\mathrm{const}$
and $\Delta_\perp Y_{lm}=-l(l+1) Y_{lm}$, we get the following differential 
equation for $F_l(t)$ after separating the variables:
\begin{equation}
\frac{1}{V^2}\,\frac{d^2 F_l(t)}{d t^2}=-\frac{l(l+1)}{a^2}\,F_l(t).
\label{eq51}
\end{equation}
This is the equation of harmonic oscillations with the cyclic frequency
\begin{equation}
\omega_l=\frac{V}{a}\,\sqrt{l(l+1)}.
\label{eq52}
\end{equation}
In particular, for the first fundamental frequency we get 
$f_1=\frac{\omega_1}{2\pi}=8-18~\mathrm{Hz}$, which is close to the
frequency of alpha rhythm \cite{39}.

From how we obtained Schumann resonances and brain waves, it should 
be clear that the similarities between them are the result of spherical 
symmetry and the small height of the ionosphere compared to the radius of 
the Earth. The existence of standing waves requires only that the material 
medium supports traveling waves that do not decay too quickly. Then the 
corresponding resonant frequencies are determined from the geometry of the 
problem and from the boundary conditions. Therefore any similarity between 
brain waves and Schumann resonances  may well be just a coincidence, and for 
their interconnection a wild stretch of our imagination will be required 
\cite{40}. Nevertheless, some arguments can be envisaged that these two 
desperately different phenomena are actually interrelated.

ELF electromagnetic fields and Schumann resonances have been present on Earth 
since the formation of the ionosphere. Therefore, they accompanied life from 
the very beginning, and it does not seem too wild to assume that in the course 
of evolution living organisms have found some useful application to these 
ubiquitous electromagnetic fields. One can even imagine that the ELF 
electromagnetic fields and related electric activity in the Precambrian 
Earth's atmosphere played the crucial role in the emergence of life according 
to the following scenario \cite{41,42}.

In the Precambrian era,  the atmosphere of the Earth was much larger and more 
similar to what Jupiter has today. In addition, the ionosphere was also much 
farther than today, about $10^3~\mathrm{km}$ far from the Earth's surface, in 
the immediate vicinity of the Van Allen belts. As a result, fluctuations of 
current in the Van Allen belts were capably of generating huge currents in 
the nearby ionosphere and the coupling of these currents to the Earth's 
metallic core would lead to an enormous and constant electrical activity. It 
is believed \cite{41}, these  electrical discharges were essential for 
production of amino acids and peptides from which the first living organisms 
were formed. This process was accompanied by an intense background of the ELF 
electromagnetic field, which could affect the formation and functionality of 
the first living cells and organisms. 

It has been suggested that these atmospheric  ELF background fields played 
a major role in the evolution of biological systems, especially in the early 
stages of evolution \cite{43}. In particular, the dominant brain wave 
frequencies may be the evolutionary result of the presence and effect of this 
ELF electromagnetic background \cite{44}. This idea is to some extent 
supported by the amazing fact that many species exhibit, irrespective of the 
size and complexity of their brain, essentially similar low-frequency 
electrical activity \cite{43}.

Various remote sensing systems of living organisms, such as visual system or 
the infrared sensors of snakes, have been developed due to the presence of 
electromagnetic energy in the corresponding parts of the spectrum. On early 
Earth, there was a significant amount of electromagnetic energy in the ELF 
portion of the spectrum. Thus, we can expect that organisms could adapt and 
somehow use this part of the electromagnetic spectrum too,  in particular the 
Schumann peaks of the Earth's  ELF electromagnetic field \cite{44}. The 
following observation provides some support for this idea.

Heat shock genes are responsible for adapting organisms to harsh environmental 
conditions. They are ubiquitous, present in various organisms from bacteria to 
humans and represent the most conservative and ancient group of genes. The
proteins encoded by these genes (heat shock proteins, HSPs) serve as molecular 
chaperones, which help in the repair, folding and assembly of nascent proteins 
during stress and prevent the accumulation of damaged cellular proteins.

It has been experimentally demonstrated that the ELF  electromagnetic fields  
can induce various heat shock proteins and, in particular, HSP70, like a real 
heat shock. The most surprising fact was that the electromagnetic fields 
caused the synthesis of HSP70 at an energy density of fourteen orders of 
magnitude lower than in heat shock \cite{45}. 

Such extraordinary sensitivity to the ELF magnetic fields (unlike ELF electric 
fields, magnetic fields easily penetrate biological tissues) should have 
a good evolutionary basis.  Astrophysical simulations show that shortly after 
the formation of the solar system, giant planets Jupiter and Saturn begin 
to migrate inward or outward. This planetary migration destabilizes the orbits 
of Neptune and Uranus into eccentric ellipses. As a result of this, the ice 
giants begin to cross the planetesimal disk beyond the orbit of Neptune and 
gravitationally scatter these planetesimals, forcing many of them to go along 
the Earth-crossing trajectories. The resulting so-called Late Heavy 
Bombardment (LHB) could have both positive and negative consequences for 
the emergence of life \cite{46}. In any case, the first living cells are 
expected to face grave dangers of powerful bombardment by meteorites 
(LHB tail). Thus, it can be assumed that the cells could use ELF magnetic 
pulses as a kind of early warning system that gives them time to prepare for 
other really dangerous stressors such as the heat pulse and the blast wave, 
which often follow the electromagnetic pulse \cite{47}.

However, a very detailed analysis in \cite {48} indicates that, from the point 
of view of the conventional classical physics, it remains a mystery that very 
weak ELF electromagnetic fields can cause any biological effect at all. The 
problem is the thermal noise. If we assume that random electric fields in 
biological tissues generated by thermal fluctuations of charge densities are 
correctly described by the Johnson-Nyquist formula, as in ordinary conductors, 
then the inevitable conclusion is that, for external ELF electric fields 
weaker than $300~\mathrm{V}/\mathrm{m}$, and for external ELF magnetic fields 
weaker than $50~\mu\mathrm{T}$,  it seems impossible to influence biological 
processes, since any effects generated by such fields in the body will be 
masked by thermal noise \cite{48}. This objection is known as the 
$\mathrm{kT}$ problem.

Despite of these categorical conclusions, biologists continued the 
experimental attempts to detect biological effects of ELF electromagnetic 
fields, remembering the words of Szent-Gyorgyi that "the biologist depends 
on the judgment of the physicist, but must be rather cautious when told that 
this or that is improbable" \cite{49}.

As a result of these attempts, a diverse and incontrovertible evidence had been
accumulated indicating that ELF radiation has important effects on the 
functioning of cells \cite{50}. We cite only a few reviews of the subject 
\cite{51,52,53,54,55,56}, where further references can be found.

The usual formulation of the  $\mathrm{kT}$ problem is based on several 
implicit assumptions that are not always justified \cite{57,58}. For example,
the Johnson-Nyquist formula assumes that the system is in a thermal 
equilibrium. Even so, although a living organism as a whole is very far 
from thermal equilibrium, the nature of this non-equilibrium is such that the 
concept of a well-defined temperature nevertheless exists \cite{59}. In fact, 
the physical nature of the biological effects of ELF fields still remains 
an enigma, and more work is needed to elucidate the comprehensive 
mechanisms behind these effects \cite{56,59}.

\section{Concluding remarks}
As we have seen, the magnitudes of the Schumann resonance fields are 
expected to increase tremendously after the Chicxulub impact. Nevertheless, 
this effect will be rather short-lived, since Schumann resonance fields decay 
rapidly due to low Q-factors of resonances.

In the long run, the impact of this magnitude will cause a very serious 
environmental damage. As a result, stratospheric  dust, sulfates released as a
result of impact, and soot from extensive worldwide forest fires caused by 
exposure to the impact related thermal radiation,  can lead to significant 
climate changes over decades  (impact winter) \cite{60} and hence modify 
lightning activity, which is the main source of energy for Schumann 
resonances.

In addition, blast wave for some time distorts the ionosphere and changes 
the frequencies of Schumann resonances. It is clear from (\ref{eq23A}) that 
not only the frequency of the first Schumann resonance will be changed, but 
all other Schumann resonances too. However, it is difficult to estimate the 
magnitude of the effect without detailed modeling of the expected impact on 
the ionosphere. Note that after high altitude  ''Starfish'' nuclear test 
explosion, all resonance frequencies abruptly dropped by about 0.5~Hz 
\cite{27}.

It has been suggested that ELF background atmospheric fields played a 
major role in the evolution of biological systems, and in particular that 
Schumann resonances are used for synchronization by living organisms
\cite {43,61}. If so, then the abrupt change in the Schumann resonance 
parameters after the Chicxulub impact could have a stressful effect,
contributing to a devastating load on the global biosphere, including 
dinosaurs.

A somewhat similar idea can be found in \cite{62}, where it was suggested
that the influence of the ELF and ultra-law frequency (ULF) electromagnetic 
fields produced by widespread earthquakes and volcanism in the dinosaur
era stimulated their growth in size,  and when these phenomena were no 
longer so common dinosaurs became extinct for a number of reasons, 
including the loss of intensity of the ULF/ELF electromagnetic fields.

In addition, we note that the possible abrupt distortion of the Schumann 
resonance parameters after a full-scale nuclear war is an additional risk 
factor that has not been studied, and, therefore, the danger that humanity 
will become extinct like dinosaurs after such a catastrophic event is perhaps 
underestimated by politicians.

\section*{Acknowledgments}
The work is supported by the Ministry of Education and Science of the Russian 
Federation. The author is grateful to the anonymous reviewer for constructive 
suggestions, which helped to improve this paper.


\end{document}